\documentclass[aps,pra,twocolumn,10pt,floatfix,amsmath,amssymb,superscriptaddress,longbibliography]{revtex4-2}
\usepackage{graphicx}
\usepackage{amsmath,amssymb}
\usepackage{booktabs}
\usepackage[hidelinks]{hyperref}
\usepackage{tikz}
\usepackage{pgfplots}
\pgfplotsset{compat=1.16}
\usetikzlibrary{arrows.meta,calc}

\begin{document}

\title{Characterization of nested Walsh parity-check filters in a single-photon eight-mode register on a cloud photonic processor}

\author{E.~Tully}
\author{J.~Washburn}
\author{M.~Simons}
\email[Corresponding author: ]{msimons@recognitionphysics.org}
\affiliation{Recognition Physics Institute, Austin, TX 78701, USA}

\date{June 2026}

\begin{abstract}
We characterize how faithfully a commercial cloud photonic
processor implements two nested Walsh parity-check filters acting
on the spatial-mode amplitudes of a single photon.  Eight modes
are indexed by the vertices of the three-cube $Q_3$; the filters
realize the classical $[8,7,2]$ single-parity-check code (the
zero-sum ``neutral'' subspace $\mathcal{N}$) and the $[8,4,4]$
extended Hamming code (the parity-checked subspace
$\mathcal{S}\subset\mathcal{N}$, with one DC and three face-parity
syndrome channels).  These are first-quantized path/mode
encodings of classical codes: nothing is corrected, and all
reported probabilities are conditional on postselected
single-photon detection---photon loss, the platform's dominant
error, is removed by the postselection and is not detected by the
checks.  On Quandela's Belenos processor, across more than
340{,}000 postselected detections, the implemented filters
reproduce the intended linear algebra to percent-level leakage
floors: neutral inputs reach residual DC-port leakage of
$0.02\%$--$1.1\%$ depending on input (mean $0.6\%$), a
$\approx21\times$ suppression relative to the ideal $0.125$
DC-capture baseline ($31.6\times$ relative to the measured
non-neutral control, the gap quantifying an on-chip calibration
bias); injected DC contamination produces a monotonic,
calibratable soft error signal; and the three face-parity
syndrome channels route to their predicted ports with $94$--$99\%$
selectivity, for which we report the full confusion matrix.
Leakage remains far below non-neutral controls after one to three
applications of a sector-preserving unitary core, with cross-depth
differences dominated by per-compilation calibration systematics
rather than gate-cycle physics.  We identify and quantify the
systematics that limit all of these figures (fixed-pattern
separator bias, $\pm 0.02$ per-point calibration offsets,
order-of-magnitude compilation scatter at the $10^{-3}$ leakage
level), and report an opportunistic Hong--Ou--Mandel degradation
episode, during which the suppression vanished and recovered with
recalibration, as a diagnostic observation.  The measurements
verify leakage suppression and syndrome routing for one
postselected single-photon block; they do not demonstrate error
correction, closed-loop recovery, feed-forward control, a
fault-tolerance threshold, or protection against photon loss.
\end{abstract}

\maketitle

\section{Introduction}

Quantum error protection requires encoding logical information
into a subspace of a larger physical Hilbert
space~\cite{nielsen2010,gottesman1997}.  The overhead of this
encoding (the ratio of logical to physical resources) is a
central figure of merit, although its precise meaning depends
on the physical register.  In multi-qubit
fault-tolerance roadmaps, the surface code~\cite{fowler2012}
operates at code rates of order a few percent at fault-tolerant
distances~\cite{litinski2019,acharya2025}, and even recent
low-overhead constructions such as bivariate bicycle
codes~\cite{bravyi2024} carry substantial qubit overhead, whereas
small error-detecting codes such as
$[\![4,2,2]\!]$~\cite{vaidman1996,gottesman1997} reach high rates
only at low distance.  These stabilizer-code rates are quantified
for context in Sec.~\ref{sec:context}; they are not directly
comparable to the single-photon spatial-mode dimension rate used
below, but they highlight a general design pressure in quantum
information experiments: the need for experimentally accessible
error-detecting subspaces with directly measurable leakage or syndrome
channels~\cite{klm2001,obrien2009,flamini2019}.

We present a single-photon construction on the three-cube
$Q_3$ (the three-dimensional hypercube graph whose eight
vertices are the $3$-bit binary strings, with edges joining
strings that differ in exactly one bit) that achieves
spatial-mode dimension rate $7/8$ for a
7-dimensional error-detecting subspace on an 8-mode register, and
test representative states from this subspace on a commercial
photonic chip using more than 340{,}000 single-photon events.  The
construction uses the geometry of the cube to define two nested subspaces: a
7-dimensional neutral subspace $\mathcal{N}$ for high-rate
suppression, and a 4-dimensional parity-checked subspace
$\mathcal{S}\subset\mathcal{N}$ with one DC/sum syndrome and
three independent face-parity syndrome channels.  Both subspaces
are defined by linear constraints: $\mathcal{N}$ is the zero-sum
hyperplane of $\mathbb{C}^8$, and $\mathcal{S}$ is the kernel of a
$4\times8$ Walsh parity-check matrix (one DC/sum row and the three
face-parity check functionals of the cube), with coding-theoretic minimum
distances $d(\mathcal{N})=2$ and $d(\mathcal{S})=4$.  As
linear codes these are the classical $[8,7,2]$
single-parity-check code and the $[8,4,4]$ first-order
Reed--Muller code $\mathrm{RM}(1,3)$ (equivalently the extended
Hamming code)~\cite{macwilliams1977,reed1954,muller1954}; our
contribution is not the codes themselves but their single-photon
spatial-mode realization on cloud photonic hardware, the
cube-derived Gray-code gate ordering, and the unitary-core
invariance test described below.  The four code
states form a $\mathbb{Z}_2^3$-character basis, and the same
construction extends to the $d$-cube $Q_d$ with $d+1$ checks at
fixed distance~4.  The encoding is
first-quantized:
states are normalized vectors in $\mathbb{C}^8$ indexed by
spatial mode, as in high-dimensional photonic encodings
based on discrete path or mode degrees of
freedom~\cite{erhard2020}.  The error weight of a vector counts
the number of perturbed mode amplitudes.  This is a qudit-style
encoding (throughout, a single-photon path/mode qudit in
dimension~8) and not a multi-qubit stabilizer code, so the
rate should be interpreted as a property of the single-photon
spatial-mode encoding.  Relative to generic high-dimensional
single-photon encodings on path or mode degrees of
freedom, the cube geometry serves here as an organizing
principle: it ties the DC check, face-parity syndrome
structure, and Gray-code gate ordering to a single combinatorial
object.  The experiments verify the resulting filter structure;
they do not isolate a hardware advantage attributable to the cube
labeling itself (Sec.~\ref{sec:discussion}).

On hardware, representative neutral-state inputs exhibit
$\approx21\times$ leakage suppression relative to the ideal
$0.125$ DC-capture baseline ($31.6\times$ relative to the measured
non-neutral control), and representative parity-checked
inputs exhibit $23.7\times$ syndrome suppression with
$94$--$99\%$ dominant-port selectivity
for the three face-parity channels.  The neutral-sector unitary
core keeps BALANCE leakage far below the non-neutral controls over
the tested depths.  In the ideal limit these suppressions are
exact by construction; what the measurements characterize is how
faithfully the processor implements the intended filters, the
leakage floors at which implementation imperfections enter, and
the systematics that set those floors.  The simulations of
Sec.~\ref{sec:sim} provide
idealized reference points rather than additional hardware
measurements.  The specific combination tested here is nested
single-photon spatial-mode subspaces with leakage suppression and
resolved face-parity syndrome channels on a programmable,
cloud-accessible photonic processor; we have not performed a
systematic survey of multiport Walsh- and parity-routing
experiments, so we state this as the configuration tested rather
than as a priority claim.  To situate it, we
contrast the present work with the two closest integrated and
all-optical baselines along the axes that define it.  Bell
\emph{et al.}~\cite{bell2014} demonstrated a graph-state
error-correction code with syndrome extraction and correction, but in
a multi-photon polarization-entangled encoding of a single logical
qubit realized on a bulk and fibre source, not on a programmable,
cloud-accessible mesh.  Zhang \emph{et al.}~\cite{zhang2023} encoded
error correction on a programmable \emph{integrated} photonic chip,
again in a multi-qubit (dual-rail) encoding.  Neither targets the
combination reported here: a single-photon spatial-mode
(path/mode-qudit) register laid out on the eight-vertex cube geometry,
carrying a high-rate neutral subspace, three simultaneously resolved
face-parity syndrome channels, and cloud-only access with no
user-side cryogenics.  The distinguishing axes are therefore the
single-photon spatial-mode encoding and the resolved face-parity
syndrome structure on a cloud-accessible programmable processor.
Generic programmable multiport interferometry on single photons is
itself well established~\cite{carolan2015,flamini2019}, and
leakage suppression and programmability \emph{per se} are shared
with this prior work; the specific element here is the simultaneous
exposure of the DC/sum channel and three independently resolved
face-parity syndrome channels on that platform.  The experiment should
therefore be read as a
single-block demonstration of leakage suppression, syndrome routing,
and neutral-sector retention in a single-photon spatial-mode
register, not as a demonstration of closed-loop recovery,
feed-forward correction, or a fault-tolerance threshold.

\section{Theory}
\label{sec:theory}

\subsection{Physical register and error model}

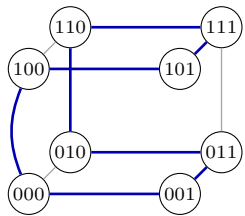
\begin{figure}[!t]
\centering
\begin{tikzpicture}[scale=1.0,
  vertex/.style={circle,draw=black,fill=white,inner sep=1pt,
                 minimum size=4mm,font=\scriptsize},
  edge/.style={gray!70, line width=0.5pt},
  graycycle/.style={blue!70!black, line width=1.0pt}]
  \node[vertex] (v000) at (0,0)        {000};
  \node[vertex] (v001) at (2.0,0)      {001};
  \node[vertex] (v011) at (2.55,0.55)  {011};
  \node[vertex] (v010) at (0.55,0.55)  {010};
  \node[vertex] (v100) at (0,1.65)     {100};
  \node[vertex] (v101) at (2.0,1.65)   {101};
  \node[vertex] (v111) at (2.55,2.2)   {111};
  \node[vertex] (v110) at (0.55,2.2)   {110};
  \draw[edge] (v010)--(v011);
  \draw[edge] (v110)--(v111);
  \draw[edge] (v010)--(v110);
  \draw[edge] (v011)--(v111);
  \draw[edge] (v000)--(v010);
  \draw[edge] (v001)--(v011);
  \draw[edge] (v100)--(v110);
  \draw[edge] (v101)--(v111);
  \draw[graycycle] (v000)--(v001);
  \draw[graycycle] (v001)--(v011);
  \draw[graycycle] (v011)--(v010);
  \draw[graycycle] (v010)--(v110);
  \draw[graycycle] (v110)--(v111);
  \draw[graycycle] (v111)--(v101);
  \draw[graycycle] (v101)--(v100);
  \draw[graycycle] (v100) to[bend right=22] (v000);
\end{tikzpicture}
\caption{Vertex layout of the 3-cube $Q_3$ used as the index
set for the eight photonic modes.  Each vertex carries a
3-bit address $b_2 b_1 b_0$.  The blue cycle is the
Gray-code Hamiltonian cycle that orders the modes for the
archived neutral-sector core.  The three pairs of opposite
faces (constant $b_0$, $b_1$, $b_2$) correspond to the three
face-parity check functionals $M_0, M_1, M_2$ of
Eq.~\eqref{eq:C}; together with the DC/sum row, these form the
four checks defining $\mathcal{S}$.}
\label{fig:cube}
\end{figure}

The state space is $\mathbb{C}^8$, with modes indexed by the
vertices of $Q_3$ via 3-bit labels $b_2 b_1 b_0$.  A single
photon occupies a normalized superposition of these 8 spatial
modes (Fig.~\ref{fig:cube}).  For a subspace
$\mathcal{C}\subset\mathbb{C}^8$, the \emph{minimum support
weight}
\[
  d(\mathcal{C}) \;=\; \min_{0 \neq x \in \mathcal{C}}
  \bigl|\mathrm{supp}(x)\bigr|
\]
is the minimum number of nonzero mode amplitudes in any
nonzero element of $\mathcal{C}$; equivalently, it is the
minimum Hamming weight of a nonzero codeword in the
first-quantized (spatial-mode) representation, i.e.\ the
standard coding-theoretic minimum distance applied to
mode-amplitude vectors.  In the additive
first-quantized path-amplitude error model used here, a
nonzero perturbation supported on fewer than $d$ modes cannot
lie entirely in $\mathcal{C}$ and is therefore detected by
projection onto the complementary syndrome space.

The error model used throughout this paper is the
\emph{first-quantized path-mode error model}: a
``single-mode amplitude perturbation'' is an additive local
change to one mode amplitude.  Concretely, a relative attenuation
or imbalance on mode $k$ acts as $v_k \to v_k + \delta$ with
$\delta\in\mathbb{C}$, and a phase rotation on an occupied mode
acts as $v_k \to v_k + (e^{i\phi}-1)\,v_k$; in either case the
perturbation occupies a single mode coordinate, so the support
weight of the error is~$1$ and the Hamming-weight (minimum-distance)
argument above applies directly.  Physically, this models
relative mode-dependent attenuation or imbalance in one
waveguide, as well as coherent leakage or contamination isolated
to a single path, conditional on the postselected single-photon
readout used in the experiments.  We adopt this single-mode channel
because per-waveguide amplitude imbalance and isolated path
contamination are the leading-order imperfections of a programmable
mesh acting on a postselected single photon; it is thus the natural
first-order error model for this register, fixed independently of
the subspace construction and not chosen to match it.  A local
phase rotation produces the same syndrome signal to first order
when the affected mode is occupied.  This is not the standard multi-qubit Pauli channel, and
the detection and syndrome guarantees we report are for the
single-photon spatial-mode channel.  Errors of support weight
greater than one (coherent multi-mode phase errors from
miscalibrated mesh elements, correlated insertion-loss patterns
across several modes, and source phase drift) fall outside this
single-mode model and are not protected by the present checks.  We
use the word
\emph{protection} throughout in this operational sense: the linear
constraints \emph{detect and suppress} such single-mode amplitude
perturbations, through projection onto the complementary syndrome
space and the associated leakage and syndrome readout.  This is
distinct from active, closed-loop quantum error correction, which
would additionally require complex-syndrome measurement and
feed-forward and is not claimed here (see Sec.~\ref{sec:limitations}).

Three features of Fig.~\ref{fig:cube} drive the construction
in the remainder of this section.  First, every vertex enters
the sum constraint $\sum_i v_i = 0$ on equal footing, so the
neutral subspace $\mathcal{N}$ is geometrically the
zero-mean hyperplane of $\mathbb{C}^8$.  Second, the three
pairs of opposite faces (constant-$b_0$, constant-$b_1$, and
constant-$b_2$) supply the three sign patterns
$(-1)^{b_j}$ that we use as face-parity check functionals below;
the cube has no other independent opposite-face partition, which
is why $\mathcal{S}$ has exactly three face-parity syndrome
channels in addition to the DC/sum channel.
Third, the blue Gray-code Hamiltonian cycle visits each vertex
once and fixes the mode ordering used by the archived
neutral-sector core of Sec.~\ref{ssec:dynamics}.  That core is the
circuit element whose algebraic action preserves the neutral
constraint and the stated checks.  Thus Fig.~\ref{fig:cube}
summarizes the common cube geometry underlying the check
functionals (sum, $M_0, M_1, M_2$), the gate layout, and the
error-detecting subspaces ($\mathcal{N}, \mathcal{S}$).

\subsection{Neutral subspace \texorpdfstring{$\mathcal{N}$}{N}: spatial-mode dimension rate \texorpdfstring{$7/8$}{7/8}, distance~2}

The neutral subspace is defined by
\[
  \mathcal{N} \;=\; \bigl\{\,v \in \mathbb{C}^8\,:\,
  \textstyle\sum_i v_i = 0\,\bigr\},
  \qquad \dim\mathcal{N} = 7.
\]
We refer to this zero-sum hyperplane as the \emph{neutral}
subspace, following the terminology used for this construction in
Ref.~\cite{washburn2026}.  Write
$\mathbf{1} = (1,1,\dots,1)^{\!\top} \in \mathbb{C}^8$ for the
all-ones vector.  The orthogonal projector onto $\mathcal{N}$ is
the BALANCE operator
\begin{equation}
\label{eq:balance}
  B \;=\; I_8 \,-\, \tfrac{1}{8}\,\mathbf{1}\,\mathbf{1}^{\dagger},
\end{equation}
in which $\mathbf{1}\,\mathbf{1}^{\dagger}$ is the rank-one outer
product (not a scalar), so $\tfrac{1}{8}\,\mathbf{1}\,
\mathbf{1}^{\dagger}$ is the orthogonal projector onto the
uniform (DC) direction $\mathbf{1}$ and $B$ removes that
component.  Any nonzero
additive single-mode perturbation in the sense above changes
the sum and is therefore detectable through the DC readout
port; for a pure local phase rotation, the signal is
proportional to the occupied amplitude in that mode.  The
minimum-weight codeword $(1,-1,0,\ldots,0)$ has weight 2, so
$d(\mathcal{N}) = 2$.  As a linear code over the mode amplitudes,
$\mathcal{N}$ is the classical $[8,7,2]$ single-parity-check
code~\cite{macwilliams1977}.  We call
$R_{\mathrm{mode}}(\mathcal{C}) = \dim(\mathcal{C})/8$ the
\emph{spatial-mode dimension rate} of a subspace
$\mathcal{C}\subset\mathbb{C}^8$; thus
$R_{\mathrm{mode}}(\mathcal{N}) = 7/8$.  This rate counts protected
dimensions per physical mode in the single-photon register and is
not a multi-qubit stabilizer-code rate.

Structurally, $\mathcal{N}$ is analogous to a decoherence-free
subspace or noiseless code: information is placed in a subspace
orthogonal to a collective mode, and the implemented gates are
chosen to leave that subspace
invariant~\cite{zanardi1997,lidar1998,kwiat2000}.
The analogy is limited, however.  Standard DFS theory protects
against noise operators with a specified collective action on the
encoded system; here the protected direction is the uniform spatial
mode of a postselected single photon, the measured soft-error signal
is the BALANCE/DC leakage produced by first-quantized mode-amplitude
perturbations, and the cube geometry supplies the face-parity
syndrome channels and Gray-code gate ordering.  Thus the present
claim is not a generic DFS claim, but an explicit spatial-mode
construction whose protection is read out by the leakage and
syndrome measurements below.

\subsection{Parity-checked subspace \texorpdfstring{$\mathcal{S}$}{S}: distance~4 with one DC/sum and three face-parity syndrome channels}

The three face-parity check functionals are represented by
$M_j = \mathrm{diag}\bigl((-1)^{b_j}\bigr)$ for
$j\in\{0,1,2\}$, i.e.\ by the row constraints
$\sum_b (-1)^{b_j} v_b = 0$.  Together with the DC/sum
constraint, they define the parity-check matrix
\begin{equation}
\label{eq:C}
C \;=\; \begin{pmatrix}
1 &  1 &  1 &  1 &  1 &  1 &  1 &  1 \\
1 & -1 &  1 & -1 &  1 & -1 &  1 & -1 \\
1 &  1 & -1 & -1 &  1 &  1 & -1 & -1 \\
1 &  1 &  1 &  1 & -1 & -1 & -1 & -1
\end{pmatrix}.
\end{equation}
The rows of $C$ are the constant function and the three
first-order Walsh (coordinate) functions on the cube, so $C$ is
the Walsh-sign (complexified) realization of the standard
parity-check matrix of the first-order Reed--Muller code
$\mathrm{RM}(1,3)$, equivalently the $[8,4,4]$ extended Hamming
code~\cite{macwilliams1977,reed1954,muller1954}: the $\pm1$
entries are the real Walsh characters, not the $\mathrm{GF}(2)$
parity-check matrix itself.  The
parity-checked subspace $\mathcal{S} = \ker(C)$ therefore has
dimension $8-4 = 4$.  Every $3$-column subset of $C$ has full rank
(verified exhaustively for all $\binom{8}{3} = 56$ triples by the
routine \texttt{all\_triples\_full\_rank} in \texttt{operators.py}
in the reproducibility package), which implies
$d(\mathcal{S}) \geq 4$; this distance-4
property is exactly the standard minimum distance of $[8,4,4]$.
The weight-4 codeword
$(1,-1,-1,1,0,0,0,0) \in \mathcal{S}$ saturates the bound, so
$d(\mathcal{S}) = 4$.

The same counting gives a useful mathematical scaling reference on
the $d$-cube $Q_d$.  With $M=2^d$ spatial modes, the neutral
subspace defined by the single DC/sum check has dimension $M-1$
and spatial-mode dimension rate $1-1/M$.  Adding the $d$
coordinate face-parity checks gives a parity-checked subspace of
dimension $M-(d+1)=M-\log_2 M-1$ and rate
$1-(d+1)/M$.  Thus the number of checks grows only logarithmically
with mode count while the same first-order Walsh-check construction
retains distance~4 for $d\geq 2$ (a weight-4 rectangle in the
hypercube saturates the bound).  This is a property of the
linear-algebraic construction, not a demonstrated multi-block
scaling architecture; in particular the logarithmic check count is
a statement about the algebra only.  The physical resource cost
scales unfavorably: the mode count $M=2^d$ and the number of modes
over which a single photon must remain coherent both grow
exponentially in $d$, while the protected distance stays fixed
at~4.

The 8 columns of $C$ are pairwise distinct, so a single-mode
amplitude perturbation $\delta$ on mode $m$ produces the unique
syndrome $s = \delta\,c_m$, where $c_m$ denotes the $m$-th
column of $C$.  This is the algebraic basis for
single-mode error identification on $\mathcal{S}$.

\subsection{Dynamics operator}
\label{ssec:dynamics}

The hardware dynamics are defined operationally from the archived
compiled circuits that were submitted to the Belenos processor.
Index the standard basis $\{e_k\}_{k\in\mathbb{Z}_8}$ by the eight
modes of Sec.~\ref{sec:theory} and recall
$\mathbf{1}=\sum_k e_k=(1,\dots,1)^{\!\top}$.  Since
$\mathcal{N}=\mathbf{1}^{\perp}$, a unitary $U$ satisfies
$U\mathcal{N}=\mathcal{N}$ if and only if $\mathbf{1}$ is an
eigenvector of $U$; this is the only gate property the
construction uses.  BALANCE is the orthogonal projector $B$ of
Eq.~\eqref{eq:balance}, which fixes $\mathcal{N}$ pointwise and
annihilates $\mathbf{1}$.

Let $S$ denote the BALANCE separator unitary whose first output
mode is the normalized uniform vector, and let $U_d$ denote the
archived compiled circuit unitary for the Experiment~4 neutral
jobs with $d=0,1,2,3$ core iterations
(\texttt{E4f\_neutral\_no}, \texttt{E4f\_neutral\_1x},
\texttt{E4f\_neutral\_2x}, and \texttt{E4f\_neutral\_3x}).  The
implemented neutral-sector core is the one-cycle operator isolated
from these payloads,
\begin{equation}
\label{eq:Rhat}
  R_{\mathcal{N}} \;\equiv\;
  S^{\dagger} U_1 U_0^{\dagger} S .
\end{equation}
The same core reconstructs the archived two- and three-cycle
circuits as
$U_d = S R_{\mathcal{N}}^{d} S^{\dagger} U_0$ to machine
precision.  Thus the hardware test of Sec.~\ref{sec:hardware}
probes repeated applications of one well-defined unitary core,
followed by one BALANCE separator readout and single-photon
postselection.

Table~\ref{tab:params} lists the numerical checks.  The full
$8\times8$ matrix and the extraction script are included in the
reproducibility package; the script regenerates the matrix from the
raw Perceval payloads and prints the corresponding eigenphases.  The
extracted core is real and orthogonal (unitary to roundoff,
$\det R_{\mathcal{N}}=-1$), fixes $\mathbf{1}$ with eigenvalue $+1$,
and therefore preserves $\mathcal{N}$.  No claim of
optimality (fastest mixing, minimal depth, or otherwise) is made for
this particular core; it is the archived representative used to
exercise the neutral sector.

\begin{table}[t]
\caption{Validation checks for the implemented neutral-sector core
$R_{\mathcal{N}}$ of Eq.~\eqref{eq:Rhat}.  Values are regenerated
from the archived Experiment~4 payloads by the extraction script in
the reproducibility package; the operator is real and orthogonal and
fixes the uniform vector with eigenvalue $+1$.}
\label{tab:params}
\begin{ruledtabular}
\begin{tabular}{ll}
Check & Value \\
\hline
Field, type & real, orthogonal \\
Unitarity residual & $1.33\times10^{-15}$ \\
Uniform-vector eigenvalue & $1.000000+0.000000i$ \\
$\det R_{\mathcal{N}}$ & $-1$ \\
Eigenphases (deg) & $0,\ \pm15.28,\ \pm43.37,$ \\
 & $\pm166.82,\ 180$ \\
Depth-2 reconstruction residual & $1.11\times10^{-15}$ \\
Depth-3 reconstruction residual & $1.50\times10^{-15}$ \\
\end{tabular}
\end{ruledtabular}
\end{table}

The projector $B$ is not a unitary mesh element: on hardware it is
realized only as the BALANCE/parity separator followed by
single-photon postselection (Sec.~\ref{sec:methods}).  Therefore the
compiled hardware circuits apply $R_{\mathcal{N}}^d$ ($d\geq0$)
and then perform a single separator readout; the projection onto
the neutral sector is an algebraic description of the ideal
readout, not an additional compiled gate at each cycle.

\section{Simulation benchmarks}
\label{sec:sim}

Eight idealized simulation benchmarks provide reference points for
the construction; the compact summary below lists all eight.  These
simulations are not hardware measurements and were not preregistered
predictions.

\begin{center}
\footnotesize
\begin{tabular}{lp{0.78\columnwidth}}
\hline
ID & Benchmark and result \\
\hline
B1 & Subspace structure: rank $= 7$, $\|B^2{-}B\| = 0$ \\
B2 & Logical action (shallow): rank $= 7$, overlap $= 1/7$ \\
B4 & BALANCE cadence: positive mean fidelity gain$^{\dagger}$ \\
B6 & Deep unitarity: $\kappa = 1.000$ at 200 cycles \\
B7 & Single-mode decoder: $100\%$ recovery (500 trials; Wilson 95\% CI $[99.2\%, 100\%]$) \\
B8 & Parity-check vs.\ BALANCE: wins $16/16$ conditions \\
\hline
\end{tabular}
\end{center}
\noindent {\footnotesize $^{\dagger}$Representative magnitude from
the archived dephasing model in \texttt{simulations.py}; this
benchmark is model-dependent, and the script reproduces the
qualitative outcome rather than the exact value (see below).}

Benchmarks B1--B6 probe the neutral subspace $\mathcal{N}$;
B7--B8 probe the parity-checked subspace $\mathcal{S}$.  The
benchmarks fall into two classes.  Benchmarks B1, B2, B6, and
the single-mode decoder of B7 are \emph{exact} algebraic checks:
fully determined by the construction, they reproduce to numerical
precision.  They verify the projector structure
($\|B^2 - B\| = 0$), the rank and mixing of the shallow gate cycle
(rank $7$, with a mean squared basis-state overlap of $1/7$
reflecting near-uniform mixing within the seven-dimensional
neutral subspace), the
deep-unitarity condition number ($\kappa = 1.000\,000$), and
$100\%$ single-mode syndrome recovery.  Benchmarks
B4, the weight-2 decoder of B7, and B8 are \emph{model-dependent}
stochastic simulations evaluated under a transparent per-mode
dephasing model with a fixed random seed; the magnitudes quoted
for them are representative values of that model rather than exact
predictions.  The archived script (\texttt{simulations.py})
reproduces their qualitative outcomes (a positive BALANCE-cadence
fidelity gain, weight-2 recovery beyond the formal distance-4
guarantee, and $P_{\mathcal{S}}$ outperforming BALANCE across all
16 conditions), and not the exact percentages, which depend on
the noise-model implementation and seed.  Two benchmarks reported
in earlier versions of this manuscript (a dimension-weighted
throughput comparison and an internal gate-count consistency
check) have been removed: the former compared a simulated channel
against a fixed reference constant on a metric that favors any
high-dimensional subspace by construction, and the latter checked
only the internal bookkeeping of the simulation code; neither
constitutes an independent benchmark, and both remain available in
the archived \texttt{simulations.py} for completeness.  We expand
below on the
three results most directly relevant to the hardware (B6, B7,
B8).  These checks inform the construction but are not directly
probed by the hardware experiments, whose results are reported in
Sec.~\ref{sec:hardware}.

\textit{B6 (deep unitarity).}  Evaluated on the ideal, noiseless
operator, the implemented core
$R_{\mathcal{N}}$ maintains condition number
$\kappa = 1.000\,000$ through 200 simulated cycles, with all
seven singular values equal to 1 to numerical precision
(benchmark~B6); the condition number and singular
values are computed from the exact propagated operator, so no
noise model enters this check.  This provides the ideal
sector-preservation reference for the Experiment~4 hardware test
of repeated applications of $R_{\mathcal{N}}$ followed by BALANCE
readout.

\textit{B7 (single-mode decoder).}  In simulation only, a
syndrome-based decoder recovers $100\%$ of single-mode amplitude
perturbations across 500 randomized trials (benchmark~B7; Wilson
95\% CI $[99.2\%, 100\%]$).  Each trial applies a
single-mode amplitude perturbation of random magnitude (drawn
uniformly from $[0,1]$, in the first-quantized error model of
Sec.~\ref{sec:theory}) on a uniformly random mode of a random code
state; the decoder assigns the syndrome to the check column most
parallel to it (routine \texttt{b7a\_single\_mode\_decoder}), which
succeeds uniquely because the eight check columns are pairwise
distinct.  A greedy sequential decoder, which iteratively assigns the
syndrome to the most parallel check column, subtracts that
column's contribution, and repeats (routine
\texttt{b7b\_weight2\_greedy}), also recovers a large fraction of
weight-2 patterns (a representative $71.4\%$ under the archived
dephasing model); this exceeds the formal distance-4 guarantee and
reflects decoder behavior rather than a sharper distance claim, and
the exact recovery fraction is model-dependent.

\textit{B8 (parity-checked vs.\ BALANCE).}  Across 16
noise-depth conditions (four per-cycle
dephasing amplitudes
$\sigma_\phi \in \{0.01, 0.03, 0.10, 0.30\}$ crossed with four
depths $\{10, 50, 100, 200\}$ cycles), the parity-check
projector $P_{\mathcal{S}}$ (the orthogonal projector onto $\mathcal{S}$)
outperforms BALANCE alone in every one of the 16 conditions, with
a representative fidelity improvement of up to ${\sim}15\%$ at the
highest noise level in benchmark~B8.  The clean sweep across all
conditions is the robust outcome; the exact improvement magnitude
is model-dependent.  This benchmark isolates the modeled
contribution of the three additional parity constraints beyond the
single sum constraint.

\section{Hardware platform and methods}
\label{sec:methods}

All experiments were performed on Quandela
Belenos~\cite{quandela_belenos,quandela_catalog}, a 24-mode
universal photonic processor accessed remotely through the
Perceval software framework~\cite{heurtel2023}, with no user-side
cryogenic infrastructure.  The experiment uses the standard
linear-optical model in which programmable interferometers
implement unitary transformations on photonic modes through
beam-splitter and phase-shifter decompositions and integrated
multiport meshes~\cite{reck1994,clements2016,klm2001,carolan2015,flamini2019}.

\textit{Mode selection.}  We used 8 logical modes of the
24-mode Belenos processor, exposed in Perceval as an 8-mode
circuit with ports $0$--$7$.  We did not manually optimize or
selectively choose a specific physical subset of modes for fidelity;
the circuit was submitted through the Belenos/Perceval compiler
and the platform handled the assignment of logical to physical
modes.  Because the assignment is performed by the compiler
rather than chosen by hand, the specific physical mode indices
(and any run-to-run reassignment) are not fixed in advance; the
logical-to-physical mapping returned by the compiler for each
job, together with the calibration timestamp for every
acquisition (and hence for each row of Tables~\ref{tab:balance}
through~\ref{tab:hom}), is recorded in the raw Perceval payloads
provided in the reproducibility package.

\textit{State preparation.}  Input states were generated in
software by QR decomposition.  For each target neutral state
$|\psi\rangle$, we constructed a unitary
$U_{\mathrm{prep}}$ with
$U_{\mathrm{prep}}\,|0\rangle = |\psi\rangle$ by placing the
normalized target in the first column of an otherwise-identity
matrix and completing it to an orthonormal basis with the
Householder QR routine \texttt{numpy.linalg.qr}, fixing the sign
convention so that $\mathrm{diag}(R)$ is real-positive and the
first column of $U_{\mathrm{prep}}$ equals $|\psi\rangle$ up to a
global phase (routine \texttt{prep\_unitary} in
\texttt{operators.py}).  The identical convention was used for
every input in Tables~\ref{tab:neutral}, \ref{tab:stab},
and~\ref{tab:parity}.  Because the completion of the remaining
seven columns is convention-dependent, the ordering of the
sub-percent residual DC components across inputs (for example the
especially small residual of the $|0\rangle - |4\rangle$ pair) is
a property of this particular preparation rather than of the
target states; the exact preparation unitaries are archived in the
reproducibility package, and a mathematically equivalent but
different QR convention would in general redistribute that
sub-percent residual leakage among inputs without altering the
two-to-three-tier separation between neutral, non-neutral control,
and pure-DC states.  We then submitted
the composite circuit
$U_{\mathrm{tot}} = U_{\mathrm{sep}}\,U_{\mathrm{prep}}$ as a
single Perceval circuit, with a single photon injected into
mode $0$ as the physical input.  Here $U_{\mathrm{sep}}$ denotes
the readout separator (BALANCE or parity separator) applied after
state preparation; both $U_{\mathrm{prep}}$ and $U_{\mathrm{sep}}$
were compiled into a single interferometer submission to the
Belenos backend.  The preparation unitary was
not separately re-fit to hardware data, and was not modified
between the calibrated, degraded, and restored chip states of
the Hong--Ou--Mandel control (Sec.~\ref{sec:hardware}).

\textit{Input states.}  In the mode basis $(e_0,\dots,e_7)$,
with $e_k$ the single-photon state in mode $k$, the inputs used
in the heralding experiments are as follows.  The
\emph{non-neutral control} is a single computational-basis mode,
$|0\rangle = e_0$ (net sum $1$, uniform overlap $1/8$, hence
expected dump probability $0.125$).  The neutral pair inputs are
$\tfrac{1}{\sqrt2}(e_0 - e_1)$ and $\tfrac{1}{\sqrt2}(e_0 - e_4)$
(a third neutral pair, $\tfrac{1}{\sqrt2}(e_0 - e_2)$, is used
only for the face-parity probe of Experiment~5B),
and the \emph{balanced} $4{+}4{-}$ input is the $(-1)^{b_2}$ sign
pattern
$\tfrac{1}{\sqrt8}(e_0{+}e_1{+}e_2{+}e_3{-}e_4{-}e_5{-}e_6{-}e_7)$;
all three satisfy $\sum_i v_i = 0$.  The four $\mathcal{S}$-basis
inputs are the $\mathbb{Z}_2^3$ characters
$|\chi_a\rangle = \tfrac{1}{\sqrt8}\sum_x (-1)^{a\cdot x} e_x$ with
$a \in \{011, 101, 110, 111\}$ (written as $a_2 a_1 a_0$;
Sec.~\ref{sec:hardware}, Experiment~5A), i.e.\ the labels
$b_0\oplus b_1$, $b_0\oplus b_2$, $b_1\oplus b_2$, and
$b_0\oplus b_1\oplus b_2$.  These inputs were chosen to probe
distinct structural features of the construction: local neutral
pairs, the self-conjugate neutral pair $|0\rangle - |4\rangle$ whose
opposite signs on the two self-conjugate modes ($k=0$ and $k=4$) yield
the smallest residual DC component, a balanced
four-plus/four-minus neutral state, and the four character basis
states spanning $\mathcal{S}$.

\textit{Operating conditions.}  The Perceval remote-processor
metadata for \texttt{qpu:belenos} reported a \texttt{Clock (MHz)}
field of $4.94$, which we adopt as the single-photon clock rate;
this device-level metadata value is distinct from public platform
shot-rate summaries.  The same device-metadata record also
reported a second-order correlation $g^{(2)}(0) = 0.019$
and a transmittance of $4.84\%$.  The Hong--Ou--Mandel
visibility~\cite{hong1987} exceeded $90\%$ on the calibrated
chip.  Each test configuration collected approximately
$10{,}000$ single-photon events.  Statistical uncertainties on
every measured fraction
$\hat p = x/n$ are reported as $95\%$ Wilson confidence
intervals~\cite{wilson1927},
$\big(\hat p + z^2/2n \pm z\sqrt{\hat p(1-\hat p)/n + z^2/4n^2}
\big)\big/(1+z^2/n)$, with $z = 1.96$.  Suppression-ratio
uncertainties are computed by the log-ratio (delta) method:
$\mathrm{SE}\big(\log R\big) =
\sqrt{(1-\hat p_a)/x_a + (1-\hat p_b)/x_b}$ for
$R = \hat p_a/\hat p_b$.

\textit{Multi-photon contamination.}  The residual multi-photon
content of the source is measured directly from the archived
exports, whose decoded port-count distributions retain every
photon-number outcome, not only the postselected
single-photon subset.  Across the $350{,}000$ events recorded in
the six experiments, $171$ registered two photons and none
registered three or more, a directly observed multi-photon
detection fraction of $0.049\%$, and at most $0.10\%$
($\le 10/10{,}000$) in any single configuration (the same
$0.049\%$ rate holds over the full set of archived campaign
exports; audited by \texttt{photon\_number\_audit.py} in the
reproducibility package).  These detected pairs are flagged and
removed by the single-photon postselection.

The only remaining
multi-photon path is a two-photon emission in which one photon is
lost before detection and the survivor is counted as a single
photon; using the reported $4.84\%$ transmittance $\eta$, the
ratio of one-detected to two-detected outcomes of a photon pair is
$2(1-\eta)/\eta \approx 39$, placing such lost-twin survivors at
$\approx 1.9\%$\footnote{The lost-twin survivor fraction is
estimated as $[2(1-\eta)/\eta]\,(n_2/N)$, where $\eta = 4.84\%$ is
the transmittance, $n_2 = 171$ is the number of detected
two-photon events, and $N = 350{,}000$ is the total number of
detections; this gives $39\times(171/350{,}000)\approx1.9\%$.} of
the postselected single-photon counts
(consistent with the source $g^{(2)}(0) = 0.019$).  A
lone surviving photon, however, traverses the same linear circuit
$U_{\mathrm{tot}}$ as the intended single photon and undergoes the
same single-particle coherent cancellation, landing with the
identical port distribution
$|\langle\mathrm{port}|U_{\mathrm{tot}}|\psi\rangle|^2$; lacking
only the two-photon interference of a genuine pair, it should
reproduce the single-photon signal rather than produce a distinct
dump or syndrome bias.  This argument relies on the surviving
photon's marginal one-mode state matching the intended
single-photon input.  Both the one- and two-photon
components of the source are injected into the same single spatial
mode (mode~$0$), so a lost-twin survivor is a mode-$0$ single
photon identical to the intended input and traverses
$U_{\mathrm{tot}}$ with the same port distribution; any residual
bias is controlled by the source's spatial-mode purity and is
bounded above by the $\le 1.9\%$ survivor weight, below the
two-to-three-order-of-magnitude leakage and syndrome separations
reported in Sec.~\ref{sec:hardware}.  The multi-photon
component is therefore either explicitly discarded (detected
pairs, $0.049\%$) or constrained to the same single-particle port
statistics up to the survivor bound; it neither
accounts for the residual $\sim 1\%$ preparation/calibration
leakage floor of Experiments~2 and~3 (Eq.~\eqref{eq:cal}) nor
alters the headline suppression ratios of
Sec.~\ref{sec:hardware}.

\begin{figure*}[!t]
\centering
\begin{tikzpicture}[
  font=\footnotesize,
  block/.style={draw=black, rounded corners=2pt, align=center,
                minimum width=2.7cm, minimum height=0.9cm,
                fill=gray!8},
  readout/.style={draw=blue!60!black, rounded corners=2pt,
                  align=center, minimum width=2.7cm,
                  minimum height=0.9cm, fill=blue!6},
  detect/.style={draw=black, rounded corners=2pt, align=center,
                 minimum width=2.7cm, minimum height=0.9cm,
                 fill=green!8},
  arrow/.style={-{Triangle[length=2.4mm,width=2.0mm]}, line width=0.7pt}
]
\node[block] (input) at (0,0) {single photon\\in mode $0$};
\node[block] (prep) at (3.2,0) {software-defined\\preparation $U_{\rm prep}$};
\node[block] (core) at (6.4,0) {optional core\\$R_{\mathcal{N}}^d$};
\node[readout] (balance) at (9.8,0.8) {BALANCE readout\\DC dump port};
\node[readout] (parity) at (9.8,-0.8) {parity separator\\DC + face syndromes};
\node[detect] (ports) at (13.1,0) {postselected\\single-photon ports};

\draw[arrow] (input) -- (prep);
\draw[arrow] (prep) -- (core);
\coordinate (coreToBalance) at ($(core.east)+(0,0.18)$);
\coordinate (coreToParity) at ($(core.east)+(0,-0.18)$);
\coordinate (portsInTop) at ($(ports.west)+(0,0.18)$);
\coordinate (portsInBottom) at ($(ports.west)+(0,-0.18)$);

\draw[arrow] (coreToBalance) -- (balance.west);
\draw[arrow] (coreToParity) -- (parity.west);
\draw[arrow] (balance.east) -- (portsInTop);
\draw[arrow] (parity.east) -- (portsInBottom);

\node[align=center, font=\scriptsize] (corelabel) at (6.4,-1.35)
{Experiment~4 only:\\depth $d=0,1,2,3$};
\draw[gray!70, line width=0.45pt] (corelabel.north) -- (core.south);
\node[align=center, font=\scriptsize, text=blue!60!black] (balancelabel) at (9.8,1.75)
{Experiments~1--4, 6\\BALANCE route};
\draw[blue!60!black, line width=0.45pt] (balancelabel.south) -- (balance.north);
\node[align=center, font=\scriptsize, text=blue!60!black] (paritylabel) at (9.8,-1.75)
{Experiment~5\\parity route};
\draw[blue!60!black, line width=0.45pt] (paritylabel.north) -- (parity.south);
\end{tikzpicture}
\caption{Experimental workflow for the single-photon hardware
tests.  A software-defined preparation unitary maps the injected
single photon to a target neutral or parity-checked input state.
Depending on the experiment, the prepared state is sent directly
to a separator or first through $d$ applications of the
neutral-sector unitary core $R_{\mathcal{N}}$.  BALANCE readout
measures the DC leakage channel, while the parity separator routes
the DC/sum and face-parity syndrome channels to output ports.  All
reported probabilities are normalized to postselected
single-photon detections across the eight logical ports.}
\label{fig:workflow}
\end{figure*}
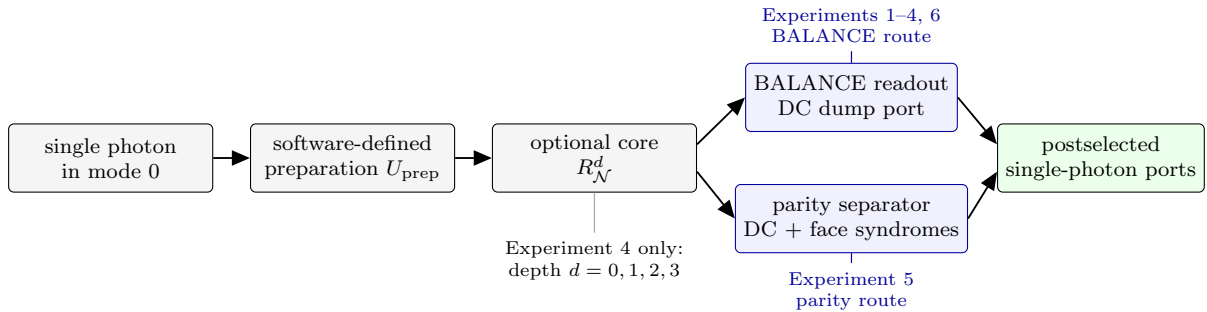

\textit{Acquisition, postselection, and backend version.}  Each
configuration was executed on the Belenos backend through the
Perceval framework during the April~2026 campaign.  The Perceval
remote-processor metadata identified the backend as
\texttt{qpu:belenos} (``Belenos QPU''), and the exported job
payloads recorded \texttt{pcvl\_version}~1.0.1; the API-reported
backend software stack was \texttt{mosaiq-belenos}~2.7.12,
\texttt{hardware-core}~3.7.4, \texttt{pcvl-worker}~1.4.0,
\texttt{perceval-quandela}~1.1.0,
\texttt{universalchipworker}~1.7.0, and \texttt{exqalibur}~1.1.1.
The
Hong--Ou--Mandel control (Sec.~\ref{sec:hardware}) spans the
calibrated, degraded, and restored chip states recorded on
April~4--6, 2026.  Circuits were compiled to the native Belenos
interferometer mesh by the default Perceval/Belenos compiler with
no manual remapping (cf.\ \emph{Mode selection}); the use of a
programmable multiport mesh follows the standard universal
interferometer model~\cite{reck1994,clements2016}.  Readout is
single-photon: we postselect on events with exactly one detected
photon across the eight output ports, discarding zero-photon
(loss) events and multi-photon and dark-count-flagged events.
The reported dump and syndrome probabilities are normalized to
the total postselected single-photon counts per configuration,
so mode-independent loss divides out and only the relative
port-occupation distribution enters.  Across all configurations
the single-photon postselection retains $99.9\%$ of detected
events (range $99.9$--$100\%$), with no systematic dependence on
input state or core depth; the small discarded fraction is the
multi-photon component audited above.  Absolute photon loss
(zero-photon events) is not registered by the platform's event
counter and, being mode-independent, divides out under this
per-configuration normalization, so all reported probabilities are
conditional on single-photon detection.  Claims involving loss or
attenuation should therefore be read as relative path-amplitude
statements conditional on this postselection, not as measurements
of absolute optical loss.  The per-configuration raw port-count
distributions, the exact preparation unitaries, and the
postselection masks are provided in the Supplemental Material
accompanying this manuscript (see the Data Availability section).
Figure~\ref{fig:workflow} summarizes the corresponding
hardware workflow.

\section{Hardware results}
\label{sec:hardware}

The six hardware experiments follow the workflow of
Fig.~\ref{fig:workflow}.  Experiments~1--4 and~6 read out the
prepared state on the BALANCE (DC dump) port, with Experiment~4
inserting $d=1$--$3$ applications of the neutral-sector core
$R_{\mathcal{N}}$ before readout; Experiment~5 instead routes the
state through the parity separator to resolve the DC/sum and three
face-parity syndrome channels.

\subsection{Experiment 1: BALANCE separator (80{,}000 events)}

Single photons in each of the 8 computational modes are sent
through the BALANCE separator.  The dump port (port~0) should
capture $1/8$ of the input from each mode.

\textit{Result.}  The dominant output port is correct for all
8 inputs (Table~\ref{tab:balance}).  The mean dump probability
across modes is $0.188 \pm 0.030$ (the spread denotes the
across-mode standard deviation; ideal value $0.125$); the
best-performing input (mode~2) gives $14.0\%$ on the dump port
(Wilson 95\% CI $[0.133, 0.147]$).  The separator routes the
dominant output as predicted, with a measurable calibration bias in
the dump probability that is carried through in the controls below.
As an implementation-fidelity figure for the separator, the
measured DC-capture channel deviates from its ideal $0.125$ value by
a root-mean-square of $0.070$ across the eight modes (a mean upward
bias of $50\%$).  This fixed-pattern bias is the dominant systematic
of the BALANCE experiments: it is the same offset that inflates the
non-neutral control in Experiment~2 and is the origin of the gap
between the directly measured $31.6\times$ and the
calibration-referred $\approx21\times$ suppression reported there.

\begin{table}[t]
\caption{Experiment 1: BALANCE separator per-input dump
probability.  A single photon is injected in each computational
mode~$m$; the dump port (port~$0$) should capture the ideal
fraction $1/8 = 0.125$.  Dump probability $\hat p$ is the port-$0$
fraction of $n$ postselected single-photon events; uncertainties are
95\% Wilson confidence intervals.  The best-performing (lowest dump)
mode is boldfaced.}
\label{tab:balance}
\begin{ruledtabular}
\begin{tabular}{lrc}
Input mode & $n$ & $\hat p$ (95\% CI) \\
\hline
$m=0$ & 9{,}993 & $0.193\,[0.185,0.201]$ \\
$m=1$ & 9{,}994 & $0.157\,[0.150,0.164]$ \\
$m=2$ & 9{,}992 & $\mathbf{0.140}\,[0.133,0.147]$ \\
$m=3$ & 9{,}990 & $0.178\,[0.171,0.186]$ \\
$m=4$ & 9{,}992 & $0.188\,[0.181,0.196]$ \\
$m=5$ & 9{,}995 & $0.237\,[0.229,0.246]$ \\
$m=6$ & 9{,}995 & $0.224\,[0.216,0.233]$ \\
$m=7$ & 9{,}991 & $0.186\,[0.179,0.194]$ \\
\end{tabular}
\end{ruledtabular}
\end{table}

\subsection{Experiment 2: Neutral-state heralding (50{,}000 events)}

Photons prepared in neutral superpositions
$\sum_i v_i = 0$ are sent through the BALANCE separator.

\begin{table*}[t]
\caption{Neutral-state heralding on Belenos.  The control row is
the non-neutral input; the uniform (pure DC) row is the all-modes-in-phase
reference; all remaining rows are neutral inputs ($\sum_i v_i = 0$).
The best-performing (lowest dump) neutral input is boldfaced.
Dump probability $\hat p$ is
the fraction of photons measured on port~$0$ out of $n$
events; uncertainties are 95\% Wilson confidence intervals.}
\label{tab:neutral}
\begin{ruledtabular}
\begin{tabular}{lrcr}
Input & $n$ & $\hat p$ (95\% CI) & Expected \\
\hline
Control (non-neutral) & 9{,}998 & $0.191\,[0.183,0.198]$ & 0.125 \\
$|0\rangle - |1\rangle$ & 9{,}998 & $0.0073\,[0.0058,0.0092]$ & 0 \\
$|0\rangle - |4\rangle$ & 9{,}995 & $\mathbf{0.0002}\,[5{\times}10^{-5},7{\times}10^{-4}]$ & 0 \\
Balanced $4{+}4{-}$ & 9{,}993 & $0.0106\,[0.0088,0.0128]$ & 0 \\
Uniform (pure DC) & 9{,}996 & $0.980\,[0.977,0.982]$ & 1.000 \\
\end{tabular}
\end{ruledtabular}
\end{table*}

\textit{Result.}  Table~\ref{tab:neutral} resolves the
expected three-tier behavior.  The non-neutral control
($\hat p = 0.191$) lies above the ideal $0.125$, consistent
with chip-level imbalance in the BALANCE separator already seen
in Experiment~1.  The pure-DC input saturates the dump port at
$0.980$, two orders of magnitude above the neutral inputs and
consistent with full DC capture up to a $\sim 2\%$
residual population in the other output ports.  The
three neutral inputs sit between these extremes in a band of
width roughly $1\%$, with the $|0\rangle - |4\rangle$ pair
delivering the lowest dump probability ($\hat p = 0.0002$);
this input places opposite signs on the two self-conjugate modes
$0$ and $4$ and, in the QR-decomposed preparation used
here, accumulates the smallest residual component on the uniform
vector.  Averaging over the three distinct neutral inputs
yields a mean of $0.6\%$ ($95\%$ CI $[0.5\%,0.7\%]$).  Referred to
the ideal DC-capture baseline $0.125$, this corresponds to a
suppression of $0.125/0.006 \approx 21\times$, which isolates the
interference suppression for neutral states against the
theoretical baseline.  The directly measured ratio relative to the
non-neutral control is larger, $31.6\times$ (log-ratio $95\%$ CI
$[27.2, 36.7]$); that control baseline $0.191$ itself carries the
fixed-pattern BALANCE calibration bias documented in Experiment~1
(mean per-mode dump $0.188$ vs.\ ideal $0.125$), so the
$31.6\times$ figure conflates genuine destructive interference on
the DC port with chip miscalibration that inflates the control
above ideal.  We report both figures, and the gap between them
quantifies the BALANCE calibration bias, not additional
physics.
The three neutral inputs themselves span $0.0002$--$0.0106$
(Table~\ref{tab:neutral}); this input-to-input range, rather than
the narrow Wilson interval on the pooled mean (which reflects
photon-counting statistics only), is the physically relevant
measure of variation, and it is dominated by how much residual
uniform component each preparation accumulates under the QR
convention of Sec.~\ref{sec:methods}.  As a between-job
reproducibility check, the $|0\rangle-|1\rangle$ neutral baseline
was acquired in three independent jobs (Experiments~2,~3, and~4)
and gave dump probabilities $0.0073$, $0.0070$, and
$0.0075$, agreeing within their Wilson intervals, so for this
shared configuration run-to-run variation is no larger than
photon-counting statistics.
Because every probability is normalized to the postselected
single-photon counts (Sec.~\ref{sec:methods}), this ratio is
independent of any mode-independent loss, which cancels between
the neutral and control inputs; the Hong--Ou--Mandel control of
Experiment~6 further indicates that the suppression is
phase-coherent rather than a normalization or intensity artifact.
The hardware therefore tests representative vectors from the
theoretically $7$-dimensional error-detecting subspace, with the
residual $\sim 1\%$ leakage consistent with preparation and/or
calibration imperfections rather than a failure of the linear
sum constraint.

\subsection{Experiment 3: Proportional error detection (70{,}000 events)}

A controlled DC contamination with strength
$c\in[0,1.6]$ is injected into a prepared neutral state and the
resulting dump probability is measured.  Concretely, the
injected state is
$|\psi(c)\rangle \propto |\psi_{\mathcal{N}}\rangle
+ c\,\hat{\mathbf{u}}$, where
$|\psi_{\mathcal{N}}\rangle\in\mathcal{N}$ is the target neutral
state and $\hat{\mathbf{u}} = \mathbf{1}/\sqrt{8}$ is the unit
uniform (DC) vector orthogonal to $\mathcal{N}$; thus $c$ is the
amplitude ratio of the injected DC component to the neutral
state.  After normalization the predicted dump (DC) probability
is the squared DC weight,
\begin{equation}
\label{eq:pc}
  p(c) \;=\; \frac{c^2}{1 + c^2},
\end{equation}
which defines the ``Theory dump'' column of
Table~\ref{tab:error} (so that $p(0)=0$ and $p(c)\to 1$ as
$c\to\infty$).

\begin{table}[t]
\caption{Error injection.  The hardware response is monotonic
in the injected DC fraction and strongly correlated with the
theoretical prediction.  Hardware uncertainties are 95\% Wilson
confidence intervals on $n \approx 10{,}000$ events per row.}
\label{tab:error}
\begin{ruledtabular}
\begin{tabular}{rrc}
$c$ & Theory dump & Hardware dump (95\% CI) \\
\hline
0.00 & 0.000 & $0.007\,[0.0055,0.0088]$ \\
0.05 & 0.003 & $0.015\,[0.0126,0.0174]$ \\
0.10 & 0.010 & $0.020\,[0.0174,0.0230]$ \\
0.20 & 0.039 & $0.038\,[0.0340,0.0414]$ \\
0.40 & 0.138 & $0.187\,[0.1800,0.1953]$ \\
0.80 & 0.390 & $0.437\,[0.4269,0.4463]$ \\
1.60 & 0.719 & $0.814\,[0.8065,0.8217]$ \\
\end{tabular}
\end{ruledtabular}
\end{table}

\textit{Result.}  Rather than summarizing the agreement by a
correlation coefficient alone, we fit the hardware response to a
two-parameter calibration model that augments the ideal curve
$p(c)$ of Eq.~\eqref{eq:pc} with a leakage floor and a
multiplicative gain,
\begin{equation}
\label{eq:cal}
  p_{\mathrm{hw}}(c) \;=\; p_{\mathrm{floor}}
  \,+\, g\,\frac{c^2}{1+c^2}.
\end{equation}
A least-squares fit to the seven points of Table~\ref{tab:error}
(two free parameters, five degrees of freedom)
gives $p_{\mathrm{floor}} = 0.010$ and $g = 1.12$: a $\sim 1\%$
preparation-leakage floor (consistent with the
$\sim 0.7\%$--$1\%$ residual of Experiment~2) and a $12\%$
multiplicative gain on a near-saturated channel.  The fit
residuals, $\hat p - p_{\mathrm{hw}}(c)$, are at the $\pm 0.02$
level (largest, $+0.024$, near $c = 0.4$) and are compatible in
sign and size with the fixed-pattern calibration offsets of the
BALANCE separator seen in Experiments~1 and~2, and not with a
departure from the $c^2/(1+c^2)$ response.  Referred to the
photon-counting (Wilson) uncertainties alone, the fit gives
$\chi^2/\mathrm{dof} \approx 24$ ($5$ degrees of freedom): the
residuals are roughly an order of magnitude larger than statistical
scatter, confirming that they are dominated by systematic
calibration offsets at the $\pm0.02$ level rather than by counting
noise or a failure of the model shape.  We therefore treat this
$\pm0.02$ residual scale as a per-point systematic budget and
regard the calibration model of Eq.~\eqref{eq:cal}, rather than a
correlation coefficient, as the operative description.  The
BALANCE separator therefore reports a continuous, monotonic, and
calibratable soft error signal across $c \in [0.05, 1.6]$, well
described over this tested range by a floor-plus-gain model with
the ideal response shape fixed.

\subsection{Experiment 4: Leakage at three compiled unitary-core depths, limited by compilation systematics (60{,}000 events)}

The neutral-sector unitary core $R_{\mathcal{N}}$
(Sec.~\ref{ssec:dynamics}) is applied $1$, $2$, and $3$ times to a
prepared neutral state, followed by BALANCE readout.

\begin{table}[t]
\caption{Neutral-sector unitary core on Belenos.  Dump probability $\hat p$ at
$0$, $1$, $2$, and $3$ iterations of $R_{\mathcal{N}}$ on a
neutral input, with two non-neutral controls (no core cycle and
$1\times$); the lowest neutral dump probability is boldfaced;
uncertainties are 95\% Wilson confidence intervals.}
\label{tab:rhat}
\begin{ruledtabular}
\begin{tabular}{lrc}
Configuration & $n$ & $\hat p$ (95\% CI) \\
\hline
Neutral, no core cycle & 9{,}995 & $0.0075\,[0.0060,0.0094]$ \\
Neutral, $1\times$ core cycle & 9{,}995 & $0.0004\,[2{\times}10^{-4},1{\times}10^{-3}]$ \\
Neutral, $2\times$ core cycle & 9{,}997 & $0.0018\,[0.0011,0.0028]$ \\
Neutral, $3\times$ core cycle & 9{,}999 & $\mathbf{0.0002}\,[5{\times}10^{-5},7{\times}10^{-4}]$ \\
Control, no core cycle & 9{,}996 & $0.196\,[0.188,0.204]$ \\
Control, $1\times$ core cycle & 9{,}998 & $0.190\,[0.182,0.197]$ \\
\end{tabular}
\end{ruledtabular}
\end{table}

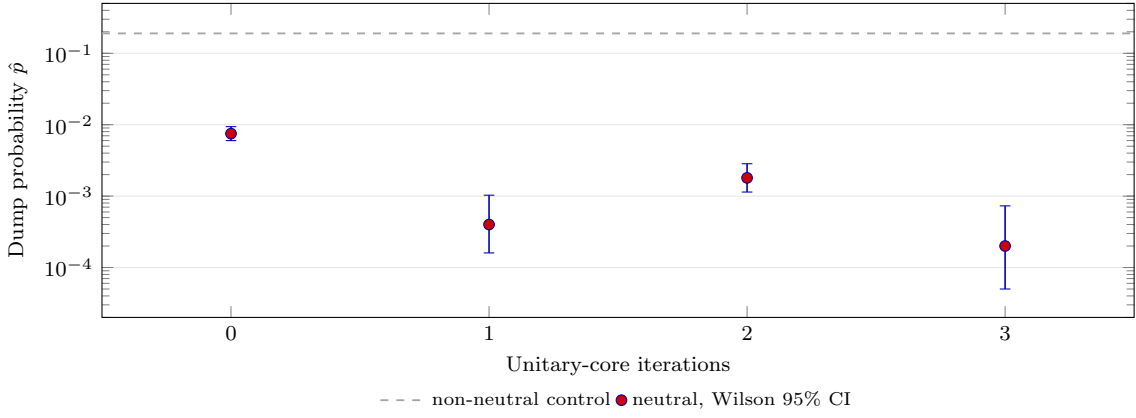
\begin{figure*}[t]
\centering
\begin{tikzpicture}
\begin{semilogyaxis}[
  width=0.85\textwidth,
  height=0.32\textwidth,
  xlabel={Unitary-core iterations},
  ylabel={Dump probability $\hat p$},
  xtick={0,1,2,3},
  xmin=-0.5, xmax=3.5,
  ymin=2e-5, ymax=5e-1,
  ymajorgrids=true,
  major grid style={gray!20},
  tick label style={font=\footnotesize},
  label style={font=\footnotesize},
  legend style={font=\scriptsize, draw=none, fill=none,
                at={(0.5,-0.20)}, anchor=north, legend columns=2},
]
\addplot[domain=-0.5:3.5, samples=2, dashed, gray!70,
         line width=0.7pt] {0.190};
\addlegendentry{non-neutral control}
\addplot+[only marks, mark=*, mark size=2pt,
         color=blue!70!black,
         error bars/.cd,
         y dir=both, y explicit,
         error mark options={rotate=90, mark size=2pt},
         error bar style={blue!70!black, line width=0.6pt}]
coordinates {
(0,0.0075) += (0,0.00189) -= (0,0.00151)
(1,0.0004) += (0,0.00063) -= (0,0.00024)
(2,0.0018) += (0,0.00104) -= (0,0.00066)
(3,0.0002) += (0,0.00053) -= (0,0.00015)
};
\addlegendentry{neutral, Wilson 95\% CI}
\end{semilogyaxis}
\end{tikzpicture}
\caption{Dump probability at the tested neutral-sector core depths
on a logarithmic scale (Experiment~4).  These are four
operating-depth points ($0$, $1$, $2$, $3$ applications) with a
non-monotonic ordering, \emph{not} a depth scan: no curve is fit
through them and the figure should not be read as a depth-scaling
trend (a dedicated $\sim$10--20 cycle scan, deferred to future
work, would be required to extract a per-cycle leakage rate).
Markers are the
hardware values; vertical bars are 95\% Wilson confidence
intervals.  The dashed line is the non-neutral control level
($\hat p = 0.190$, the $1\times$ control; the no-core-cycle
control agrees at $0.196$, statistically indistinguishable from
it, so the core does not by itself reduce the control leakage).
All four neutral points remain well
below the control band; the $1\times$ and $3\times$ points
have overlapping 95\% confidence intervals, while the
$2\times$ point lies above this lower band but well below the
no-cycle baseline.}
\label{fig:exp4}
\end{figure*}

\textit{Result.}  Table~\ref{tab:rhat} lists the numerical values;
Fig.~\ref{fig:exp4} displays the same data on a logarithmic scale.  The no-cycle neutral
baseline is $\hat p = 0.0075$; after one and three core cycles
the measured dump probabilities fall to $0.0004$ and $0.0002$,
respectively, and their Wilson intervals overlap.  The
$2\times$ point, $0.0018$, is also below the no-cycle baseline
but lies above the $1\times$ and $3\times$ lower-leakage band,
so the depth dependence is non-monotonic rather than a simple
shot-noise fluctuation.  Each depth was submitted as a different
compiled unitary, so cross-depth comparisons also include
depth-dependent compiler and calibration variance; the claim is
therefore only that leakage remains low and shows no monotonic
growth over these tested depths.  Indeed, because
$R_{\mathcal{N}}$ fixes the uniform (DC) vector $\mathbf{1}$ with
eigenvalue $+1$ (Sec.~\ref{ssec:dynamics}), the ideal core preserves the
DC/dump component of its input exactly and cannot by itself reduce
the dump leakage; in the noiseless circuit the dump probability
would be independent of depth.  The observed scatter across
depths (including the $1\times$ and $3\times$ points lying below
the $d=0$ value) is therefore consistent with differing preparation
and separator calibration for each separately compiled circuit,
not any cleaning action of the core.  The log-scale representation in
Fig.~\ref{fig:exp4} makes the main conclusion visible: all
cycled neutral points remain far below the two non-neutral
controls (no core cycle, $\hat p = 0.196$; $1\times$,
$\hat p = 0.190$).  The two
controls agree within their Wilson intervals, indicating that
the unitary core does not by itself change the non-neutral leakage
level.  The direct ratio of the $1\times$ control to the
$3\times$ neutral point is $0.190/0.0002 \approx 950$, but the
small denominator (2 events out of 9{,}999) gives a wide
log-ratio $95\%$ CI of $[2.4\times 10^{2}, 3.8\times 10^{3}]$;
we therefore treat this ratio as descriptive rather than as the
primary statistical summary.

In conventional gate-based quantum-error-correction
experiments, repeated gate application tends to degrade
fidelity~\cite{fowler2012,acharya2025}.  The present data make
a narrower claim: BALANCE leakage remains far below the
non-neutral control after one, two, and three applications of the
neutral-sector unitary core.  This agrees with the modeled
unitarity and preservation of $\mathcal{N}$ (Sec.~\ref{sec:theory},
B6 in Sec.~\ref{sec:sim}), while the non-monotonic $2\times$
point indicates that residual
preparation leakage and chip calibration also affect the
measured dump probability.  Three depths are not
sufficient to fit a per-cycle leakage rate or to separate genuine
algebraic sector preservation from cumulative coherent error
masked by the small dynamic range; accordingly, we claim only
that leakage stays well below the non-neutral control for the
three tested depths and do \emph{not} claim monotonic
sector preservation under iteration.  A dedicated depth scan to
$\sim$10--20 cycles would be required to extract a per-cycle
leakage rate, and is left to future work.  The chip realization of the core
carries an unmodeled implementation error budget
(insertion-loss imbalance, beamsplitter angle errors,
phase-calibration drift), so the observed leakage suppression
is bounded by the hardware-level fidelity of the implemented unitary,
not by the algebraic identity alone.  Simulation benchmark~B6
provides the ideal reference for sector preservation; the hardware
leakage data are compatible with that qualitative expectation
within the tested depths.  This measurement shows retention in the neutral sector as
read out by BALANCE; it does not by itself constitute a process-tomographic
measurement of the logical action within $\mathcal{N}$.

\subsection{Experiment 5: Parity-checked subspace and syndrome channels (80{,}000 events)}

The parity-check separator $U_S$ is the $\mathbb{Z}_2^3$ character
table normalized by $1/\sqrt{8}$, with rows ordered as
$a \in \{000,001,010,100,011,101,110,111\}$ (the four check
characters first, then the four code characters), so that ports
$0$--$3$ correspond to the syndrome subspace and ports $4$--$7$ to
the code subspace $\mathcal{S}$.  The port assignment follows directly from the row
structure of $C$: the four rows of $C$ correspond to the characters
with labels $a \in \{000, 001, 010, 100\}$ (the DC/sum row and the
three face-parity rows), so any state in $\ker(C) = \mathcal{S}$
has zero projection onto these four directions and is routed entirely
to ports $4$--$7$; a parity violation produces a nonzero projection
onto one or more of ports $0$--$3$.  Concretely, each basis state of $\mathcal{S}$ is
a $\mathbb{Z}_2^3$ character vector: for a label
$a = (a_2,a_1,a_0)\in\mathbb{Z}_2^3$ it is
$|\chi_a\rangle = \tfrac{1}{\sqrt{8}}\sum_{x\in\mathbb{Z}_2^3}
(-1)^{a\cdot x}\,e_x$, where $x = b_2 b_1 b_0$ runs over the
eight vertices and $a\cdot x = \bigoplus_j a_j x_j$.  The four
code states tabulated below carry the labels $b_0\oplus b_1$,
$b_0\oplus b_2$, $b_1\oplus b_2$, and $b_0\oplus b_1\oplus b_2$
(equivalently $a = 011, 101, 110, 111$ written as
$a_2 a_1 a_0$), the four characters orthogonal to the row space
of $C$.  For example,
\[
\begin{aligned}
  |b_0\oplus b_1\rangle = \tfrac{1}{\sqrt{8}}\bigl(
  &e_{000}-e_{001}-e_{010}+e_{011}\\
  &+e_{100}-e_{101}-e_{110}+e_{111}\bigr).
\end{aligned}
\]

\textit{5A. Code-space heralding.}  The four $\mathcal{S}$-basis
inputs are prepared and sent through the parity separator; port
fractions are recorded to test how well each basis state is
confined to code-space ports $4$--$7$.

\begin{table}[t]
\caption{Parity-checked-subspace heralding.  Syndrome probability
$\hat p_{\mathrm{syn}}$ is the total fraction on ports
$0$--$3$ (ideal: $0$); Port is the predicted dominant code-space output
port (ports $4$--$7$ carry the code subspace); Fid.\ is the
fraction of all detections on that predicted port;
uncertainties are 95\% Wilson
confidence intervals on $n\approx 10{,}000$ events per row.}
\label{tab:stab}
\scriptsize
\begin{ruledtabular}
\begin{tabular}{lcrr}
Basis state & $\hat p_{\mathrm{syn}}$ (95\% CI) & Port & Fid.\ \\
\hline
$b_0 \oplus b_1$ & $0.009\,[0.0073,0.0111]$ & 4 & 98.8\% \\
$b_0 \oplus b_2$ & $0.062\,[0.057,0.067]$ & 5 & 91.0\% \\
$b_1 \oplus b_2$ & $0.011\,[0.0090,0.0130]$ & 6 & 96.9\% \\
$b_0 \oplus b_1 \oplus b_2$ & $0.012\,[0.0096,0.0138]$ & 7 & 93.6\% \\
\hline
Mean ($n=39{,}982$) & $0.0232\,[0.0218,0.0248]$ & & 95.1\% \\
\end{tabular}
\end{ruledtabular}
\end{table}

Table~\ref{tab:stab} resolves the four basis-state behaviors
of $\mathcal{S}$ along two axes: total syndrome leakage
($\hat p_{\mathrm{syn}}$, ports $0$--$3$) and code-side port
fidelity (the dominant entry on its predicted code port).
Three of the four basis states cluster near
$\hat p_{\mathrm{syn}} \in [0.009, 0.012]$; the
$b_0\oplus b_2$ basis state is a clear outlier at
$\hat p_{\mathrm{syn}} = 0.062$, and its code-side port
fidelity ($91.0\%$) is correspondingly the lowest of the four.
We attribute this $5$--$7\times$ spread, tentatively, to
state-dependent calibration asymmetry rather than a global loss
of the parity-check structure: the other three basis states
retain percent-level syndrome leakage, so the parity-check routing
is intact.  This attribution is, however, a post hoc hypothesis
that the present single run cannot confirm.  To make it
quantitative and testable, we record the calibration model that
links per-mode imperfections to state-dependent syndrome leakage.
Let $D = \mathrm{diag}(1+\varepsilon_x)$ describe the residual
per-mode multiplicative error of the parity separator, where
$\varepsilon_x\in\mathbb{C}$ collects insertion-loss imbalance and
small beamsplitter-angle and phase offsets on mode $x$.  To first
order in $\varepsilon$, the amplitude that an ideal code character
$|\chi_a\rangle$ ($a\in\{011,101,110,111\}$) leaks onto check
row~$r$ (itself the character $\chi_r$ with
$r\in\{000,001,010,100\}$, the DC/sum and three face-parity
rows) is
$\langle\chi_r|D|\chi_a\rangle = \delta_{ra} +
\hat\varepsilon(a\oplus r)$, where
$\hat\varepsilon(w) = \tfrac18\sum_x (-1)^{w\cdot x}\varepsilon_x$
is the Walsh--Hadamard transform of the error profile.  Because no
code label $a$ is a check row, the predicted syndrome leakage is
\begin{equation}
\label{eq:syncal}
  p_{\mathrm{syn}}(a) \;\approx\;
  \sum_{r\in\{000,001,010,100\}}
  \bigl|\hat\varepsilon(a\oplus r)\bigr|^2,
\end{equation}
the squared Walsh weight of the calibration profile at the
frequency $a$ and its three single-face-flip neighbors.  The state
dependence is therefore fixed entirely by the Walsh spectrum of
the per-mode error: for the observed outlier $a = 101$
($b_0\oplus b_2$) the relevant frequency set is
$\{101,100,111,001\}$, so a calibration profile whose spectral
weight concentrates in this set can elevate that one character
while leaving the other three near the percent-level floor, as in
Table~\ref{tab:stab}.  Subsequent repeat acquisitions after
recalibration did move the code-state leakage floors, but not in a
session-independent way (see the June-campaign note below).  Thus
Eq.~\eqref{eq:syncal} should be read as a calibration-sensitivity
model for the observed state dependence, not as an established
diagnosis of a fixed per-mode error spectrum.  Averaged over all four inputs ($n = 39{,}982$), the
mean syndrome leakage is $2.32\%$ ($95\%$ CI
$[2.18\%, 2.48\%]$), so representative basis states of the
4-dimensional parity-checked subspace are reproduced on
hardware with percent-level residual syndrome leakage.

\textit{5B. Selective parity detection.}

\begin{table}[t]
\caption{Selective parity detection.  Each face-parity violation
activates the corresponding syndrome port; $S_x$, $S_y$, $S_z$ label
the input violating face-parity check $M_0$, $M_1$, $M_2$
respectively; selectivity is the
fraction of syndrome-subspace detections (ports $0$--$3$) routed
to the expected port, with 95\% Wilson confidence intervals.}
\label{tab:parity}
\scriptsize
\begin{ruledtabular}
\begin{tabular}{lrrc}
Input & Exp.\ port & Meas.\ port & Selectivity (95\% CI) \\
\hline
$|0\rangle - |1\rangle$ ($S_x$) & 1 & 1 & $0.985\,[0.980,0.988]$ \\
$|0\rangle - |2\rangle$ ($S_y$) & 2 & 2 & $0.944\,[0.935,0.952]$ \\
$|0\rangle - |4\rangle$ ($S_z$) & 3 & 3 & $0.942\,[0.934,0.949]$ \\
\end{tabular}
\end{ruledtabular}
\end{table}

\begin{table}[t]
\caption{Syndrome confusion matrix for Experiment~5B: the fraction
of syndrome-subspace detections (ports $0$--$3$) routed to each
syndrome port, for the three single-parity-violating inputs.
Diagonal entries (boldface) are the dominant-port selectivities of
Table~\ref{tab:parity}; off-diagonal entries are cross-talk.  Rows
sum to $1$ up to rounding.}
\label{tab:confusion}
\begin{ruledtabular}
\begin{tabular}{lcccc}
Input & Port 0 & Port 1 & Port 2 & Port 3 \\
\hline
$S_x$ $(|0\rangle{-}|1\rangle)$ & $0.008$ & $\mathbf{0.985}$ & $0.007$ & $0.001$ \\
$S_y$ $(|0\rangle{-}|2\rangle)$ & $0.008$ & $0.021$ & $\mathbf{0.944}$ & $0.028$ \\
$S_z$ $(|0\rangle{-}|4\rangle)$ & $0.015$ & $0.023$ & $0.020$ & $\mathbf{0.942}$ \\
\end{tabular}
\end{ruledtabular}
\end{table}

The three inputs are neutral pairs that each violate a single
face-parity check; the labels $S_x$, $S_y$, and $S_z$ denote
violations of $M_0$, $M_1$, and $M_2$, routed to syndrome ports
$1$, $2$, and $3$, respectively.  Each row of
Table~\ref{tab:parity} reports the
theoretically predicted syndrome port and the empirically
measured dominant syndrome port.
For all three single-parity-violating inputs the predicted and
measured ports coincide.  The dominant-port
selectivity point estimates are 94--99\%: the $S_x$ selectivity
($0.985$, 95\% CI $[0.980, 0.988]$) is higher than the $S_y$ and
$S_z$ estimates ($0.944$ and $0.942$; 95\% CIs $[0.935, 0.952]$ and
$[0.934, 0.949]$), while the
$S_y$ and $S_z$ channels are indistinguishable from each other
at this sample size.  The full syndrome confusion matrix (the
distribution of each input over the four syndrome ports
$0$--$3$) is given in Table~\ref{tab:confusion}.  Off-target
routing is below $1\%$ for $S_x$ and at the $2$--$3\%$ level for
$S_y$ and $S_z$, with no single off-target port exceeding $2.8\%$;
these off-diagonal rates bound the syndrome-label error that would
propagate into any future feed-forward correction.  The complete
per-input distributions over all eight output ports are tabulated
in the reproducibility package.  We nonetheless interpret the
measured selectivity here as syndrome-label resolution, not as a
complete syndrome-state tomography.  The three measured dominant
ports resolve the three face-parity syndrome labels.

\textit{5C. Suppression ratio.}  The mean code-space syndrome
leakage of $2.32\%$ ($95\%$ CI $[2.18\%, 2.48\%]$) contrasts
with $55.0\%$ ($95\%$ CI $[54.0\%, 56.0\%]$) for a
single computational-mode control sent through the same parity
separator (ideal syndrome fraction $1/2$).  Referred to the ideal
control fraction, the suppression is $0.5/0.0232 \approx
21.6\times$; the directly measured on-chip ratio is
$23.7\times$ (log-ratio 95\% CI $[22.2, 25.3]$), the gap
reflecting the same kind of separator calibration bias quantified
in Experiments~1 and~2 (measured control $55.0\%$ vs.\ ideal
$50\%$).  As in
Experiment~2, this ratio compares postselected single-photon
fractions, so mode-independent loss cancels
(Sec.~\ref{sec:methods}).  Because $\mathcal{S}\subset\mathcal{N}$,
the coherence sensitivity diagnosed in Experiment~6 is also
relevant to the parity-checked subspace.

The three resolved parity channels are necessary, but not
sufficient, hardware primitives for the theoretical single-mode
decoder of benchmark~B7 (100\% recovery in simulation); full
recovery also requires complex-syndrome readout and
feed-forward control.

\subsection{Experiment 6: Hong--Ou--Mandel degradation episode (diagnostic observation)}

A control experiment exploits a Belenos
recalibration/maintenance window during which the
Hong--Ou--Mandel visibility dropped from $90.7\%$ to $21.2\%$.
The degradation was opportunistic, not deliberately induced,
so the control is best interpreted as a diagnostic change in
indistinguishability rather than as an isolated knob.
We re-ran the same neutral-heralding circuit and the same
software-defined preparation in the degraded chip state and
again after the visibility was restored on recalibration.

\begin{table}[t]
\caption{Hong--Ou--Mandel control.  Identical circuit and
preparation, varying photon indistinguishability; calibration
dates are in 2026.  Dump probabilities are 95\% Wilson confidence
intervals; the calibrated state used $n\approx 10{,}000$
single-photon events and the degraded and restored states used
$n=5{,}000$.}
\label{tab:hom}
\begin{ruledtabular}
\begin{tabular}{lrc}
Chip state & HOM vis.\ & Dump prob.\ (95\% CI) \\
\hline
Calibrated (Apr.~4) & $90.7\%$ & $0.0073\,[0.0058,0.0092]$ \\
Degraded (Apr.~5--6) & $21.2\%$ & $0.481\,[0.467,0.495]$ \\
Restored (Apr.~6) & $\sim 91\%$ & $0.0058\,[0.0040,0.0083]$ \\
\end{tabular}
\end{ruledtabular}
\end{table}

Table~\ref{tab:hom} compares the dump probability across
three chip states with the circuit, the input preparation,
and the photon-counting infrastructure held fixed.  The main
recorded diagnostic change is the two-photon visibility, which
fell from $90.7\%$ to $21.2\%$ during the recalibration window
and was restored to approximately $91\%$ on completion; the
post-recalibration visibility was taken from the platform's
standard calibration report rather than separately remeasured
under our circuit, hence the ``$\sim$'' in the restored entry.
Across the
three chip states the dump probability tracks the platform
Hong--Ou--Mandel/calibration diagnostic: at calibrated visibility it sits at
$0.0073$ (95\% CI $[0.0058, 0.0092]$),
matching the corresponding row of Table~\ref{tab:neutral}; at
degraded visibility it jumps by a factor of $\sim 66$ to
$0.481$ (95\% CI $[0.467, 0.495]$); on restoration it falls
back to $0.0058$ (95\% CI $[0.0040, 0.0083]$).  The Wilson
intervals of the calibrated and restored states overlap, so
the calibrated and restored suppression levels are statistically
indistinguishable.  The degraded value of $0.481$ lies nearly two orders of magnitude
away from both the calibrated and restored values, consistent with
loss of the phase-sensitive cancellation that defines
$\mathcal{N}$ once the BALANCE separator acts on an incoherent or
poorly phased input; we do not attempt a quantitative model of its
specific magnitude, which depends on the detailed degraded chip
state (prepared-state and separator phase errors during the
maintenance window) rather than on a simple fully-incoherent
mixture.
The control is therefore consistent with, but does not by itself
isolate, a phase-coherent interference mechanism, and argues
against a purely intensity-based artifact as the source of the
calibrated suppression; a deliberate visibility scan (adding a
known temporal delay between photons to vary the indistinguishability
through a controlled range and back) would be required to isolate
the mechanism, and is left to future work.  Since $\mathcal{S}\subset\mathcal{N}$,
the parity-checked subspace requires the same coherent neutral-state preparation
and is expected to exhibit the same sensitivity to indistinguishability loss.  Because the
maintenance-window change was not an isolated control parameter,
we interpret the HOM-visibility comparison as a diagnostic of the
calibrated versus degraded photonic operating state, not as a
standalone measurement of the microscopic mechanism.

\subsection{Hardware summary}

\begin{table}[!htbp]
\caption{Summary of hardware results.}
\label{tab:summary}
\begin{ruledtabular}
\setlength{\tabcolsep}{1.4pt}%
\begin{tabular}{llr}
Experiment & Result & Events \\
\hline
1.\ BALANCE separator & 8/8 correct ports & 80{,}000 \\
2.\ Neutral heralding & $\approx21\times$ ideal ($31.6\times$ raw) & 50{,}000 \\
3.\ Error detection & calibrated soft response & 70{,}000 \\
4.\ Gate cycle & low leakage, depths 1--3 & 60{,}000 \\
5.\ Syndrome channels & $23.7\times$, 3/3 ports & 80{,}000 \\
6.\ HOM control & interference dependence & 20{,}000 \\
\hline
Total & & $> 340{,}000$ \\
\end{tabular}
\end{ruledtabular}
\end{table}

Table~\ref{tab:summary} collects the six experiments and their
event counts (the Experiment~6 entry is the exact sum of the
$10{,}000$ calibrated, $5{,}000$ degraded, and $5{,}000$ restored
acquisitions of Table~\ref{tab:hom}).  The per-configuration counts sum to roughly
$360{,}000$ single-photon acquisitions; a few baseline and control
rows are shared across experiments (for instance, the calibrated
Hong--Ou--Mandel row and the zero-contamination point of
Experiment~3 coincide with the Experiment~2 neutral baseline), so
we quote the conservative total of more than $340{,}000$ distinct
events.  Together these measurements test the operational
pieces used in the construction: BALANCE readout, leakage
suppression for representative neutral states, a calibrated soft
response to DC contamination, low leakage after repeated applications of the
neutral-sector unitary core, and resolved dominant ports for the
three face-parity syndromes.  The simulations of Sec.~\ref{sec:sim}
remain idealized references for these behaviors rather than
hardware measurements of throughput or decoder recovery.

\textit{June revision-campaign note.}  After the April data set
tabulated above, a June~2026 revision campaign repeated the
parity-checked code-state acquisitions over five Belenos calibration
sessions, scanned the archived unitary core to depths
$d=0,\ldots,15$, and ran randomized orthogonality controls.  These
follow-up data are not folded into the April event totals or the
headline ratios in Table~\ref{tab:summary}.  They support the
cautious interpretation used here: the depth scan remained below
non-neutral controls but showed non-monotonic calibration scatter,
randomized dark-port controls showed that DC-port suppression is a
generic orthogonality effect rather than a uniquely cube-specific
signature, and the parity-checked leakage floors were
session-dependent (the $b_0\oplus b_2$ code-state syndrome leakage
ranged from $3.8\%$ to $12.4\%$ across the June sessions).  The
robust hardware claim is therefore the April demonstration of
postselected leakage suppression and syndrome routing, with
calibration-dependent leakage floors, not a session-independent
suppression constant.

\section{Context and scope}
\label{sec:context}

The encoding here is a single-photon spatial-mode code rather
than a multi-qubit stabilizer code, so any comparison of code
rate is necessarily indirect.  We list the relevant numbers for
context only.  The rate axes are different and the figures below
should not be compared directly: they are given solely to situate
the construction among near-term options, not to rank it against
stabilizer codes, whose rates count logical qubits per physical
qubit against arbitrary Pauli errors.  Concretely, comparing $7/8$
with a stabilizer-code rate is not a unit conversion but a category
mismatch: the former counts protected spatial dimensions carried by
a single postselected photon, whereas the latter counts logical
qubits per encoded physical qubit against arbitrary Pauli errors.

\textit{Spatial-mode dimension rate.}  The neutral subspace
operates at rate $7/8$ ($87.5\%$) in the single-photon
spatial-mode sense: $7$ protected dimensions in $8$ physical
modes.  Multi-qubit stabilizer codes report rates on a different
resource axis (logical qubits per physical qubit against
arbitrary Pauli errors), and comparing the two is a category
mismatch rather than a unit conversion; we therefore do not
tabulate stabilizer-code rates against the spatial-mode figure
and refer the reader to
Refs.~\cite{fowler2012,acharya2025,bravyi2024,vaidman1996,gottesman1997}
for representative values.  The relevant point is only that a
high-rate single-photon spatial-mode construction with
experimentally accessible syndrome signals is of interest on
near-term photonic processors, where multi-qubit stabilizer
infrastructure is not yet available.

\textit{Behavior under repeated gates.}  In stabilizer-code
experiments, logical error per cycle is positive and overall
fidelity decreases with circuit depth~\cite{fowler2012,acharya2025}.
The Experiment~4 observation that leakage remains low under
repeated application of $R_{\mathcal{N}}$ has its ideal reference
in the unitarity of $R_{\mathcal{N}}$ established in
Sec.~\ref{ssec:dynamics} (benchmark~B6 of Sec.~\ref{sec:sim}); we do
not claim to have observed sub-unity logical error rates in the
multi-qubit sense.

\textit{Operating conditions.}  The cloud accessibility of the
processor, including the absence of any user-side cryogenic
requirement, is a property of the photonic platform and is not
unique to the present construction; the relevant point is that the
single-photon spatial-mode protection tested here does not require
custom hardware on the user side.

\textit{Relation to bosonic and loss-tolerant photonic codes.}  The
construction here protects a spatial-mode amplitude channel and is
conditional on single-photon detection; it does not address photon
loss, which is the dominant error of photonic hardware and is
removed here by postselection (Sec.~\ref{sec:methods}).
Complementary photonic strategies target exactly this channel:
bosonic codes such as the Gottesman--Kitaev--Preskill
encoding~\cite{gkp2001} protect against loss and small
displacements within a single mode, and loss-tolerant photonic
encodings~\cite{varnava2006} are designed to detect and recover
lost photons.  These operate on a different resource and error axis
from the single-photon spatial-mode parity checks demonstrated
here; a register combining spatial-mode parity filtering with
loss-tolerant or bosonic protection is a natural direction beyond
the present single-block scope.

\section{Discussion}
\label{sec:discussion}

\subsection{What the hardware demonstrates}

The hardware experiments of Sec.~\ref{sec:hardware} test the
single-block ingredients of the construction rather than a full
error-correcting architecture.  Their common feature is the
$Q_3$ layout: the DC/sum check, the soft DC error signal
(Experiment~3), the archived neutral-sector core, and the three face-parity checks
are all defined on the same eight-vertex register.  This shared
geometry is the main structural distinction from a generic
path-encoded qudit and is the feature most likely to connect this
work to adjacent high-dimensional and graph-based photonic encodings.
We emphasize, however, that the present experiments do not isolate a
hardware advantage attributable specifically to the cube
\emph{labeling}: the same circuits could be described with the modes
relabeled $0$--$7$, and the geometry functions as an organizing
principle (tying the DC check, the three face-parity syndrome
channels, and the Gray-code gate ordering to one combinatorial
object), not as an independently benchmarked physical
resource.  What the hardware shows is that this organized set of
nested checks can be prepared, applied, and read out together
on a single commercial chip.  Equivalently, the leakage- and
syndrome-suppression ratios are a measurement of how faithfully the
chip implements the intended subspace-projecting unitaries, for
which the suppression is exact by construction in the ideal limit.

\subsection{Limitations}
\label{sec:limitations}

The present work demonstrates single-block primitives of protected
spatial-mode subspaces, syndrome channels, and a
coherence-preserving unitary core.  It does not demonstrate
(i) closed-loop error
recovery using the syndrome readout, (ii) feed-forward
classical control, (iii) a multi-block tiling architecture, or
(iv) a fault-tolerance threshold.

Closed-loop recovery in particular requires measuring the
complex syndrome rather than only intensity in the
syndrome ports.  Standard photonic routes are homodyne or
heterodyne mode-overlap detection against calibrated local
oscillators, or an interferometric phase-reference
measurement that recovers both magnitude and phase of the
residual.  Our current Belenos experiments measure output
intensities and probabilities only, so they test
syndrome-channel selectivity and leakage suppression, but not
full feed-forward recovery.  Implementing complex-syndrome
readout and a feed-forward correction loop on this register
is the natural next step.

\section{Conclusion}

We have characterized, with over $340{,}000$ postselected
single-photon detection events on a commercial cloud-accessible
photonic processor, how faithfully nested Walsh parity-check
filters are implemented in a single-photon spatial-mode register:
percent-level leakage floors ($0.02\%$--$1.1\%$ depending on
input) for the zero-sum neutral subspace, a monotonic and
calibratable soft response to injected DC contamination, resolved
face-parity syndrome routing with a measured confusion matrix,
and leakage that remains far below non-neutral controls at the
three compiled depths of a sector-preserving unitary core.  The
underlying $[8,7,2]$ and $[8,4,4]$ codes are classical, the
suppressions are exact by construction in the ideal limit, and
the measured floors are set by identifiable implementation
systematics: a fixed-pattern separator calibration bias,
$\pm 0.02$ per-point offsets in the soft-error response, and
per-compilation scatter that dominates cross-depth comparisons at
the $10^{-3}$ leakage level.  All figures are conditional on
postselected single-photon detection; photon loss is not detected
by these checks.

Several interpretations remain open hypotheses rather than
established conclusions.  A subsequent June~2026 revision campaign
performed three of the targeted robustness checks: repeat
acquisitions after recalibration, a depth scan through
$d=15$, and randomized dark-port controls.  The results support the
cautious framing of this manuscript rather than strengthening the
headline claims: parity-checked leakage is calibration- and
session-dependent, the longer depth scan shows no simple monotonic
per-cycle leakage law, and dark-port suppression is not unique to
the cube labeling.  The remaining validation program is therefore a
more controlled calibration-spectrum analysis of Eq.~\eqref{eq:syncal},
a deliberate visibility scan, extension to higher $d$-cube
registers, and complex-syndrome readout with feed-forward control.
Because all reported probabilities are conditional on
single-photon postselection, a separate loss-aware implementation,
or integration with loss-tolerant photonic/bosonic protection,
would be required to turn these filters into protection against the
dominant photonic error channel.

\begin{acknowledgments}
Experiments were performed on Quandela's cloud-accessible
Belenos processor through the Perceval software framework.
The dynamics core is defined by the archived compiled circuits and
the extraction procedure reported in Sec.~\ref{ssec:dynamics}.  The
experimental claims in this manuscript rest on the explicit subspace
definitions, circuit primitives, data, and simulation benchmarks
reported here.  This
research received no external funding.

\textit{Author contributions.}
E.~T.: conceptualization, data curation, formal analysis (lead),
investigation, methodology, software, validation, visualization,
writing -- original draft, and writing -- review and editing.
J.~W.: conceptualization, investigation, methodology, software, and
supervision.
M.~S.: formal analysis, project administration, validation,
visualization, and writing -- review and editing.

The authors declare no competing financial interests.
\end{acknowledgments}

\section*{Data Availability}
The complete reproducibility package is included as Supplemental
Material with this manuscript.  It is organized as
an archival package with a README, manifest files mapping table rows
to job exports, raw Quandela/Perceval payload and result JSON files
in a dedicated raw-data directory, processed count tables in a
separate generated-output directory, backend metadata, and Python
analysis scripts for decoding the raw exports, statistical
estimation, operator algebra, simulation benchmarks, validation, and
figure/table generation.  In particular, the script
\texttt{verify\_package.py} recomputes every tabulated dump and
syndrome probability and the headline suppression ratios directly
from the archived per-port counts, resolving each manuscript row to
its job through the row-to-job manifest, and all stochastic
benchmarks use the fixed random seed \texttt{20260404}.  The package
contains no absolute local
paths, API tokens, credentials, or live internal URLs, and it does
not require live Quandela access to reproduce the reported tables
and figures from the archived raw counts.

\end{document}